\documentstyle[12pt]{article}
  \advance\oddsidemargin  by -1in
  \advance\evensidemargin by -1in
\def\lsim{\mathrel{\rlap{\lower4pt\hbox{\hskip1pt$\sim$}}
    \raise1pt\hbox{$<$}}}         
\def\gsim{\mathrel{\rlap{\lower4pt\hbox{\hskip1pt$\sim$}}
    \raise1pt\hbox{$>$}}}         

\def\be{\begin{equation}}
\def\ee{\end{equation}}
\def\br{\begin{eqnarray}}
\def\er{\end{eqnarray}}

\begin{document}
\pagestyle{empty}
\hfill{\large DFTT 39/93}

\hfill{\large July 1993}

\vspace{2.0cm}

\begin{center}

{\large \bf NON UNIVERSALITY OF STRUCTURE FUNCTIONS\\
AND MEASUREMENT OF THE STRANGE SEA DENSITY
\vspace{1.0cm}\\}
{\large V.~Barone$^{a}$, M.~Genovese$^{a}$ ,
N.N.~Nikolaev$^{b,c}$,\\
E.~Predazzi$^{a}$ and B.G.~Zakharov$^{b}$ \vspace{1.0cm} \\}
{\it
$^{a}$Dipartimento di
Fisica Teorica, Universit\`a di Torino\\
and INFN, Sezione di
Torino,    10125 Torino, Italy\medskip\\
$^{b}$L. D. Landau Institute for Theoretical Physics, \\
GSP-1,
117940,
                Moscow V-334, Russia \medskip \\
$^{c}$IKP (Theorie), KFA J{\"u}lich,
5170 J{\"u}lich, Germany}
                \vspace{1.0cm}\\

{\large \bf Abstract \bigskip\\ }

\end{center}
We show that there is no real conflict between the two
determinations of the strange sea density from the opposite--sign
dimuon production and from the difference
of the $F_2$ structure functions measured in neutrino and
muon deep inelastic
scattering.
Once non universal sea parton densities are introduced, which take
into account the effects of different mass thresholds and different
longitudinal contributions, the discrepancy is shown to disappear and
both sets of data are simultaneously
well reproduced.
No need for a large strange sea content of the nucleon emerges.

\pagebreak
\pagestyle{plain}

In a series of previous papers \cite{BGNPZ1,BGNPZ2,BGNPZ3}
we pointed out that
the sea parton densities measured in deep inelastic
scattering (DIS) are not universal: neutrinos and muons
do not probe the same strange and charm distributions. This is
the consequence of
two effects which have their origin in the dynamical
mechanism of excitation of the sea.

First of all, in
the $W$--gluon fusion process (Fig.~1c,d).
\be
W^{+} g \rightarrow c \bar s
\,\,,
\label{1}
\ee
involved in charged current neutrino DIS
the excitation of strangeness is inseparable from the
simultaneous production of charm. Notice also
that (\ref{1}) describes two processes: excitation of $c$ on $s$
and excitation of $\bar s$ on $\bar c$.
Therefore in $\nu$ DIS
we should not expect $c(x) \ll s(x)$ since we are rather probing
a charm--strange density $cs(x)$ not coinciding
with the strange density measured in $\mu$ DIS.
The mass threshold of the reaction (\ref{1}) is different from that
of the photon--gluon fusion process $\gamma^* g \rightarrow
c \bar c, s \bar s$ (Fig.~1a,b). By definition, the Bjorken
variable is related to $Q^2$ and to the mass $m_X$ of the hadronic
final state through $x = Q^2/(Q^2 + m_X^2 - m_N^2)$ and $m_X$
is such that
$m_X^2 \ge {\cal M}^2$, where ${\cal M}$ is the mass
of the excited quark--antiquark pair (to be more
precise, the sum of the masses of the lightest meson
and baryon containing the antiquark and the quark, respectively).
For strangeness and charm production, one has
${\cal M}^2 \sim 4 m_s^2 (4 m_c^2) \sim
1 (10)\, GeV^2/c^2$ in the photon--gluon fusion
and  ${\cal M}^2
\sim (m_c + m_s)^2 \sim 4 \, GeV^2/c^2$
in the $W$--gluon fusion. The contribution of an heavy flavour
to the structure function vanishes at
\be
x > x_{\rm max} = \frac{Q^2}{Q^2 + {\cal M}^2 - m_N^2}\,\,,
\label{1b}
\ee
and one can immediately see that having
${\cal M}^2 \sim (m_s + m_c)^2$ instead of ${\cal M}^2 \sim
4 m_s^2$ makes a non
negligible difference up to moderate $Q^2$ \cite{BGNPZ1}
(for a recent related discussion see also \cite{DL}).

The second source of non universality comes from the non conservation
of the vector and axial $c \bar s, s \bar c$ currents. This leads
to a large value of $R = \sigma_L/\sigma_T \sim 4 m_c^2/Q^2$
in the neutrino
excitation of charm and strangeness at low and moderately
large $Q^2$ \cite{BGNPZ3}.

The mass threshold effect and the large $R$ effect point to
opposite directions: the former depletes the transverse part of
$cs(x)$ with respect to $s(x)$ while the latter produces a relevant
longitudinal contribution to $cs(x)$. The purpose of this letter is
to show that the residual difference between $cs(x)$ and $s(x)$
can easily explain the discrepancy which presently seems to exist
between two different experimental determinations of the strange density.

To start with, let us summarize the current status of knowledge
about the strange distribution.

According to the conventional parton decomposition of the
DIS structure functions \cite{books}, in principle one can extract $s(x)$
from the difference between $F_2^{\nu}$ (measured in $\nu$ DIS)
and $F_2^{\mu}$ (measured in $\mu$ DIS) for an isoscalar target
\be
\frac{5}{6} F_2^{\nu}(x) - 3 F_2^{\mu}(x) \simeq x s(x)\,\,,
\label{2}
\ee
where we have made
the customary assumption $s^{\nu}(x) = s^{\mu}(x) = s(x)$
and $c^{\nu}(x) = c^{\mu}(x) = c(x)$, and
used $c(x) \ll s(x)$.

On the other hand, one can extract directly $s(x)$ from the data
on the opposite--sign dimuon production in $\nu N$ interaction
\cite{Brock}. In the conventional parton model,
the cross section for this process reads
\be
\frac{{\rm d}^2 \sigma(\nu N \rightarrow \mu^{+} \mu^{-} X)}{{\rm d}x
{\rm d}y} \sim 2 x s(x) \vert V_{cs} \vert^2 + x \left [ u(x)
+ d(x) \right ] \vert V_{cd} \vert^2\,\,.
\label{3}
\ee
and has been recently measured to a good accuracy by the CCFR
collaboration \cite{CCFR}.

Although the available information on the difference (\ref{2}),
coming from the CCFR measurement \cite{CCFR2} of $F_2^{\nu}$ and
from the NMC measurement \cite{NMC} of $F_2^{\mu}$, is rather poor
due to large errors, and somehow unsafe since one has to subtract
data from two different experiments, it seems clear
that the strange distribution obtained from (\ref{2})
is considerably larger than the one extracted from (\ref{3}) (see
Figs.~2,3).

 All the available global fits to the deep inelastic data are
unable to solve this puzzle. For instance, the CTEQ1M parametrization
\cite{CTEQ} reproduces the $\nu - \mu$ difference at the price of a very
large strange sea content $\kappa = 2S/ (\bar U + \bar D) =0.9$
 ($S = \int {\rm d}x \, x s(x)$, etc.) and its strange distribution
considerably overshoots the dimuon data. Martin, Roberts and Stirling
\cite{MRS1,MRS2} constrain $\kappa$ to $0.5$ and are forced to
content themselves with a fit which represents
only a reasonable compromise between the two sets of data, still
largely overshooting the dimuon results.

How the above mentioned non universality effects solve this
seemingly contradictory situation ?

Introducing the
charm--strange  distribution $cs(x)$ probed in neutrino DIS (from
now on $c(x)$ and $s(x)$ will refer to muon DIS), we obtain for
the $\nu - \mu$ difference the following decomposition
\be
\frac{5}{6} F_2^{\nu}(x) - 3 F_2^{\mu}(x) =
\frac{10}{3} x \,cs(x) - \frac{2}{3} x s(x) - \frac{8}{3} x c(x)\,\,,
\label{4}
\ee
which replaces Eq.~(\ref{2}).
The main difference between (\ref{4}) and (\ref{2}) is that the
charm--strange distribution $cs(x)$, as defined in terms of the
underlying QCD subprocess (\ref{1}), simultaneously describes the
$sc$ and the $\bar c \bar s$ excitation processes.

The ratio
\be
r(x) = \frac{2 \,cs(x)}{c(x) + s(x)}
\label{5}
\ee
is an $x$--dependent quantity, different in general from the naive
assumption $1$, which
would correspond to the conventional parton model and
to Eq.~(\ref{3}).

In the model of Ref.~\cite{BGNPZ1,BGNPZ3}
we found that at $Q^2 = 10 \, GeV^2/c^2$ the ratio $r(x)$
varies from $r(x) \simeq 1.5$ at $x = 10^{-3}$ to
$r(x) \simeq 0.8$ at $x = 10^{-1}$ and is
empirically expressible in the form
\be
r(x) = 0.72 x^{-0.13} (1-x)^{0.35}\,\,.
\label{6}
\ee
This ratio has been obtained with the choice ${\cal M}^2 = (m_c +
m_s)^2 = 4\,
GeV^2/c^2$ and can be taken as a realistic quantitative
estimate of the non universality of the strange sea density.

In terms of $r(x)$ Eq.~(\ref{4}) can be rewritten as
\be
\frac{5}{6} F_2^{\nu}(x) - 3 F_2^{\mu}(x) \simeq
\frac{1}{3} x s(x) \left [ 5 r(x) - 2 \right ]\,\,,
\label{7}
\ee
and in the small--$x$ region, $x \lsim (0.5 - 1) \cdot 10^{-1}$,
where $r(x) \gsim 1$, the $\nu - \mu$ difference
turns out to be larger than the corresponding quantity
calculated with an universal $s(x)$ distribution. This is
precisely what emerges from the comparison of the data with the
MRS($D_0'$) parametrization \cite{MRS2}.

Turning to the dimuon production, the Cabibbo unsuppressed
contribution to the cross section now reads
\be
\frac{{\rm d}^2 \sigma(\nu N \rightarrow \mu^{+} \mu^{-} X)}{{\rm d}x
{\rm d}y} \sim 2 x \,cs(x) \vert V_{cs} \vert^2
\sim x \,r(x) s(x)
\,\,,
\label{8}
\ee
and for $x \lsim 10^{-1}$ is more than a factor $2$ smaller than
the quantity predicted with $r(x) = 1$. Again, this is exactly
the discrepancy existing between the standard fits and the data.

By means of Eqs.~(\ref{4}--\ref{8}) one can now easily reproduce the
experimental determinations. To this purpose, we used for
$s(x)$ and $c(x)$ the distributions computed within our model
\cite{BGNPZ1,BGNPZ3,BGNPZ4}. The results are presented in Figs.~2,3 and show
that a very satisfactory agreement with both sets of data has been achieved.
For completeness, in Fig.~3 we present also our
predictions for the transverse contribution
to $x cs(x)$ and for the strange distribution probed by muons.

Our complete set of parton distributions gives for the
strange sea fraction at
$Q^2 = 10 \, GeV^2/c^2$ the following values
\be
\kappa^{\mu} = \frac{2S}{\bar U + \bar D} = 0.45
\,,\;\;\; \kappa^{\nu}= \frac{2 \,CS}{\bar U + \bar D} = 0.30\,.
\ee
In particular
$\kappa^{\nu}$ is found to be rather stable in the range
$10 \, GeV^2/c^2 \le Q^2 \le 30 \, GeV^2/c^2$, a feature
confirmed by the CCFR experiments, whose finding is
$\kappa^{\nu}_{\rm exp} (Q^2 = 22.2 \, GeV^2/c^2) = 0.373^{+0.048}_{-0.041}
\pm 0.018$ with no appreciable $Q^2$ variation. The reason for such
a $Q^2$ independence is that the rise with $Q^2$ of the transverse
component of $cs(x)$ is totally compensated by the simultaneous
decrease of the longitudinal contribution.

Another relevant quantity is the strange sea content
$\eta = 2 \,CS/(U + D)$ for which we found $\eta =0.07$ to
be compared to the CCFR result
$\eta_{\rm exp} = 0.064^{+0.008}_{-0.007}\pm 0.002$.

Let us now make some comments about the slow--rescaling procedure
\cite{slow}. This
is based on the assumption that in the $W^{+} + s \rightarrow c$
transition the $s$ quark, which carries momentum $\xi p_N$, is on
shell and massless, whereas the mass of the $c$ quark is retained.
Then $(\xi p_N + q)^2 = k_c^2 = m_c^2$ implies $\xi = x(1+ m_c^2/Q^2)$.
The parton distributions are rewritten as functions of $\xi$ and the
Callan--Gross relation is assumed to hold in terms of $\xi$,
so that an extra
kinematical factor $(1-y + xy/\xi)$ appears in the cross sections.
However, putting both quarks on mass shell and taking $k_s^2 = 0$
does not find any justification
in the dynamical mechanism for the excitation
of the sea, which is produced in the $W^{+}$--gluon fusion process.
The approach we presented in \cite{BGNPZ1,BGNPZ4}, based on the concept
of Fock states of the virtual photon
and of the $W,Z$ bosons, takes correctly into account the
mass effects arising from the presence of heavy quarks and cannot
be simply reduced to the slow rescaling. In a subsequent paper
we shall present our prediction for the energy dependence
of the opposite--sign dimuon rate, which is usually
regarded to be a successful validity test
of the slow--rescaling procedure.

In conclusion, we outline our results. There is no puzzling
conflict between the $\nu - \mu$ and the dimuon data. Simply, they
represent different quantities whose parton content is given by
Eqs.~(\ref{4}) and (\ref{8}). Both data sets are simultaneously
reproduced by relying on non universal sea parton densities.
The strange sea probed by neutrinos
is smaller than the one probed by muons. Their ratio $r(x)$ could be
used as a further input in the global parametrizations of the
deep inelastic structure functions and we expect, for instance, that
allowing for $r(x) \ne 1$ would considerably improve the agreement of
the MRS($D_0'$) fit with the data. Finally, there is no
need for a large intrinsic strange sea content
of the nucleon. The data seem to
favour the values $\kappa^{\mu}= 0.4-0.5$ and $\kappa^{\nu}=0.3-0.4$
at $Q^2 \simeq 20 \, GeV^2/c^2$.

\vspace{3cm}
{\large \bf Acknowledgment.} An useful correspondence with A.~Bazarko
 (CCFR collaboration) is gratefully acknowledged.

\pagebreak

\pagebreak

\begin{center}

{\large \bf Figure captions}

\end{center}

\vspace{1cm}

\begin{itemize}

\item[Fig.~1]
Photon--gluon fusion (a,b) and $W$--gluon fusion (c,d) processes
producing strange and charmed sea.

\item[Fig.~2]
Our prediction for the difference $5/6 F_2^{\nu} - 3 F_2^{\mu}$
at $Q^2 = 10 \, GeV^2/c^2$ compared to the data (CCFR for
$F_2^{\nu N}$ and NMC for $F_2^{\mu D}$).

\item[Fig.~3]
The CCFR data on the strange sea density extracted from opposite--sign
dimuon production compared to our result for $x cs(x)$ at $Q^2 =
10 \, GeV^2/c^2$ (solid line). Also shown are our predictions for the
transverse component of $x \,cs(x)$ (dotted line) and for $x s(x)$ (dashed
line).

\end{itemize}

\end{document}